\documentclass[aps,prb,twocolumn,showpacs,superscriptaddress,longbibliography]{revtex4-1}  % for review and submission
\usepackage{graphicx}  % needed for figures
\usepackage{dcolumn}   % needed for some tables
\usepackage{bm}        % for math
\usepackage{amssymb}   % for math
\usepackage{comment}
\usepackage{amsmath,accents}
\usepackage{array}
\usepackage{calc}
\usepackage{nth}

\newcommand{\etal}{\textit{et al}.}

\newcommand{\3}{$^{3+}$}

\newcommand{\YVO}{YVO$_4$}

\newcommand{\level}[3]{$^{#1}${#2}$_{#3}$}

\newcommand{\myto}{$\leftrightarrow$}

\newcommand{\Gh}{$\Gamma_{\textrm{h}}$}

\newcommand{\bs}{\boldsymbol}

\newcommand{\Caltech}{Kavli Nanoscience Institute and Thomas J. Watson, Sr., Laboratory of Applied Physics, \\ California Institute of Technology, Pasadena, California 91125, USA.}
\newcommand{\IQIM}{Institute for Quantum Information and Matter, \\ California Institute of Technology, Pasadena, California 91125, USA.}

\RequirePackage[normalem]{ulem} %DIF PREAMBLE
\RequirePackage{color}\definecolor{RED}{rgb}{1,0,0}\definecolor{BLUE}{rgb}{0,0,1} %DIF PREAMBLE
\begin{document}
% the following line is for submission, including submission to the arXiv!!
%\hspace{5.2in} \mbox{Fermilab-Pub-04/xxx-E}

\title{Controlling rare-earth ions in a nanophotonic resonator using the ac Stark shift}
\author{John G. Bartholomew}\affiliation{\Caltech}\affiliation{\IQIM}
\author{Tian Zhong}\affiliation{\Caltech}\affiliation{\IQIM}
\author{Jonathan M. Kindem}\affiliation{\Caltech}\affiliation{\IQIM}
\author{Raymond Lopez-Rios}\affiliation{\Caltech}\affiliation{\IQIM}
\author{Jake Rochman}\affiliation{\Caltech}\affiliation{\IQIM}
\author{Ioana Craiciu}\affiliation{\Caltech}\affiliation{\IQIM}
\author{Evan Miyazono}\affiliation{\Caltech}\affiliation{\IQIM}
\author{Andrei Faraon}\email{Email address: faraon@caltech.edu}\affiliation{\Caltech}\affiliation{\IQIM}
\date{\today}

\begin{abstract}
On-chip nanophotonic cavities will advance quantum information science and measurement because they enable efficient interaction between photons and long-lived solid-state spins, such as those associated with rare-earth ions in crystals. The enhanced photon-ion interaction creates new opportunities for all-optical control using the ac Stark shift. Toward this end, we characterize the ac Stark interaction between off-resonant optical fields and Nd\3-ion dopants in a photonic crystal resonator fabricated from yttrium orthovanadate (\YVO). Using photon echo techniques, at a detuning of 160~MHz we measure a maximum ac Stark shift of 2$\pi\times12.3$ MHz per intra-cavity photon, which is large compared to both the homogeneous linewidth (\Gh = 100 kHz) and characteristic width of isolated spectral features created through optical pumping ($\Gamma_{f} \approx 3$~MHz). The photon-ion interaction strength in the device is sufficiently large to control the frequency and phase of the ions for quantum information processing applications. In particular, we discuss and assess the use of the cavity enhanced ac Stark shift to realize all-optical quantum memory and detection protocols. Our results establish the ac Stark shift as a powerful added control in rare-earth ion quantum technologies. 
\end{abstract}

\maketitle

Efficient interfaces between photons and spins in solids are one foundation on which to build integrable and scalable quantum technologies for computing, communication, and metrology. One promising system for realizing photon-spin interfaces to create, control, and store quantum states is crystals containing rare-earth ions (REIs). Experiments in REI crystals have demonstrated entangled photon-pair generation~\cite{Ferguson2016, Kutluer2017, Laplane2017}, quantum memories for light~\cite{Hedges2010, Gundogan2015, Laplane2016}, and qubit operations~\cite{Longdell2004a}. These results, combined with some of the longest optical and spin coherence times in the solid state~\cite{Equall1994, Bottger2009, Zhong2015, Rancic2017} establish the future potential of REI quantum technologies. 

In most cases, quantum optical protocols performed in REI materials rely on large ensembles ($10^9$ ions) to compensate for the weakly allowed $4f \leftrightarrow 4f$ optical transitions~\cite{Reid2005}. Although this approach has proved effective, the use of large ensembles in doped crystals sets a macroscopic lower bound on the device size. This is because increasing the spectral-spatial density of REI dopants increases ion-ion interactions that cause added inhomogeneity and decoherence. The size restriction imposed by the use of large ensembles places limits on the integration and scalability of the REI platform. Thus, there is significant impetus to develop other methods to increase photon-ion interactions~\cite{Saglamyurek2014, Marzban2015, TZhong2015, Saglamyurek2016, Corrielli2016}. One solution is to fabricate photonic crystal resonators directly in REI crystals~\cite{TZhong2015, Zhong2016, Miyazono2016, Zhong2017, Zhong2017a}. A large increase in photon-ion coupling is achieved through cavity enhancement of the optical transition~\cite{Afzelius2010a} and strong mode confinement~\cite{Hastings-Simon2006, Sinclair2010, Saglamyurek2011, Marzban2015, Sinclair2017}. Given that previous work has shown that the optical properties of the ion ensembles are preserved within such nanophotonic devices~\cite{TZhong2015, Zhong2017a}, the platform is a significant opportunity for ensemble and single REI technologies. Furthermore, these photonic crystal resonators are suited to harnessing phenomena that are more commonly associated with systems with strong optical transitions, such as the ac Stark shift (ACSS).

\begin{figure*}[t]
  \begin{center}
\includegraphics[width=\textwidth, trim = 0.2cm 17.2cm 10cm 0.5cm, clip = true]{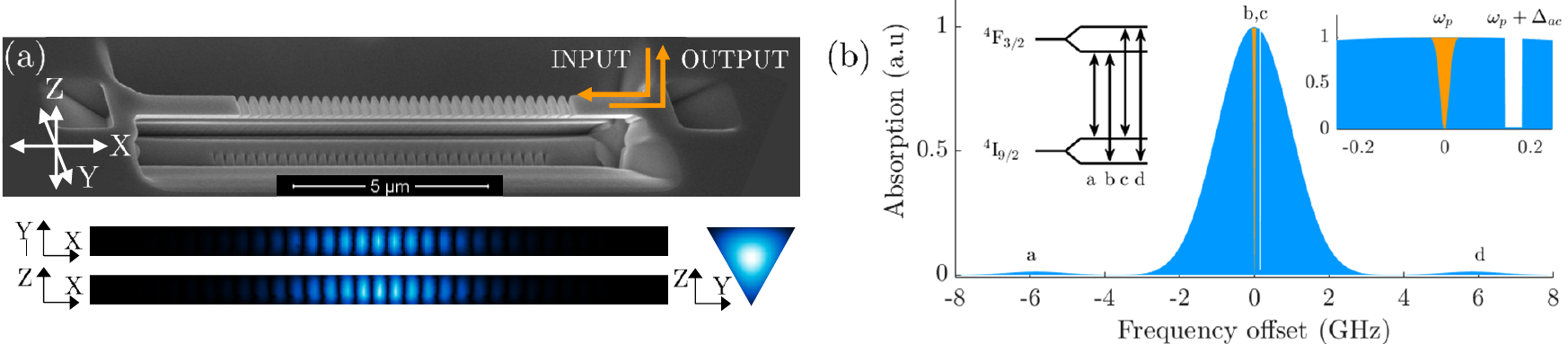}
  \end{center}
\caption{(a) Photonic crystal cavity fabricated on the surface of a Nd\3:\YVO\ substrate. The cavity is one-sided allowing measurements to be performed in the reflection mode. Below the scanning electron microscope images are cross-sections of $|E_z|$ in the cavity, which illustrate the spatial inhomogeneity of the cavity field. (b) The optical absorption of the \level{4}{I}{9/2}(Z1) \myto\ \level{4}{F}{3/2}(Y1) transition in Nd\3:\YVO\ modeled from the spin Hamiltonian, which shows the spectral subset of ions that contribute to the photon echo signal (at frequency $\omega_p$), and the transmission trench prepared for the ACSS pulses (at frequency $\omega_p + \Delta_{ac}$). The inset on the left shows the energy level structure of the transition, and the inset on the right shows the spectral region of interest in greater detail.}
\label{fig::DeviceAndSpectrum}
\end{figure*} 

Strong photon-ion interactions, including a strong single photon ACSS, offer an important additional degree of control for REIs. In this work we characterize the ACSS in an on-chip nanophotonic resonator containing Nd\3\ ions. The ACSS has been investigated previously in bulk REI crystals~\cite{Chaneliere2015, McDonald2016}, where the interaction resulted from $\sim10^{12}$ photons interacting with $\sim10^{9}$ ions. We build on this work in a different regime: where the strength of the ACSS is sufficiently large to allow the study of the interaction between a single photon with approximately $4\times10^3$ ions. The ACSS was probed using a photon echo technique, which allowed the measurement of the maximum ACSS in the cavity and the inhomogeneity of the interaction across the ensemble of ions. These measurements are then used to analyze the usefulness of the ACSS as a tool in the REI quantum optics tool box. In particular, we discuss using the large ACSS to realize all-optical quantum memories based on the hybrid photon echo rephasing (HYPER) protocol~\cite{McAuslan2011a} and the atomic frequency comb~\cite{Afzelius2010}, and cross phase modulation using the protocol suggested in Reference~\onlinecite{Sinclair2016}. Our study demonstrates that the enhanced photon-ion interactions resulting from coupling of REIs to photonic crystal resonators offer new avenues for quantum technologies in these materials.

The material chosen for this work was \YVO\ doped with a Nd\3-ion impurity at a nominal level of 50~ppm (Gamdan Optics). We use the 879.9~nm transition between the lowest crystal field components of the \level{4}{I}{9/2} and \level{4}{F}{3/2} multiplets. Both these levels are Kramers doublets, the degeneracy of which is lifted in an applied magnetic field resulting in four optical transitions (see Figure~\ref{fig::DeviceAndSpectrum}(b)). This transition has been characterized previously~\cite{Sun2002, Hastings-Simon2008a, Afzelius2010} and shown to possess narrow inhomogeneous linewidths, and the largest documented optical dipole moment for REI transitions suitable for quantum memory applications~\cite{McAuslan2009}. Furthermore, optical pumping of Nd\3:\YVO\ allows the electron spin to be highly polarized~\cite{Hastings-Simon2008a, Afzelius2010, Zhong2017a}, because of long lived spin states. Notably, the Nd\3\ site has D$_{\textrm{2d}}$ symmetry and hence, has a vanishing dc Stark shift. As a result, despite the high absorption possible in this material, its use in quantum memory applications has been limited because electric fields cannot be used to control the ions. 

A one-sided photonic crystal cavity was milled on the Nd\3:\YVO\ surface perpendicular to the crystal c-axis using a focused ion beam. The nanophotonic cavity is based on a triangular nanobeam~\cite{TZhong2015, Zhong2016} and is described in more detail in Reference~\onlinecite{Zhong2017a}. To couple the TM mode illustrated in Figure~\ref{fig::DeviceAndSpectrum}(a) to the Nd\3\ optical transition, the cavity is frequency tuned through nitrogen gas condensation. The device was cooled to approximately 500~mK in a $^3$He cryostat to reduce transition broadening due to phonon interactions. In addition, a constant magnetic field of approximately 340~mT was applied at a small angle from the \YVO\ c-axis to reduce broadening from Nd\3-Nd\3\ magnetic dipole interactions. Further details about the cavity and the experimental setup are provided in the Supplemental Material.

\begin{figure}[t]
  \begin{center}
\includegraphics[width=\columnwidth, trim = 0cm 0.8cm 3.5cm 1cm, clip = true]{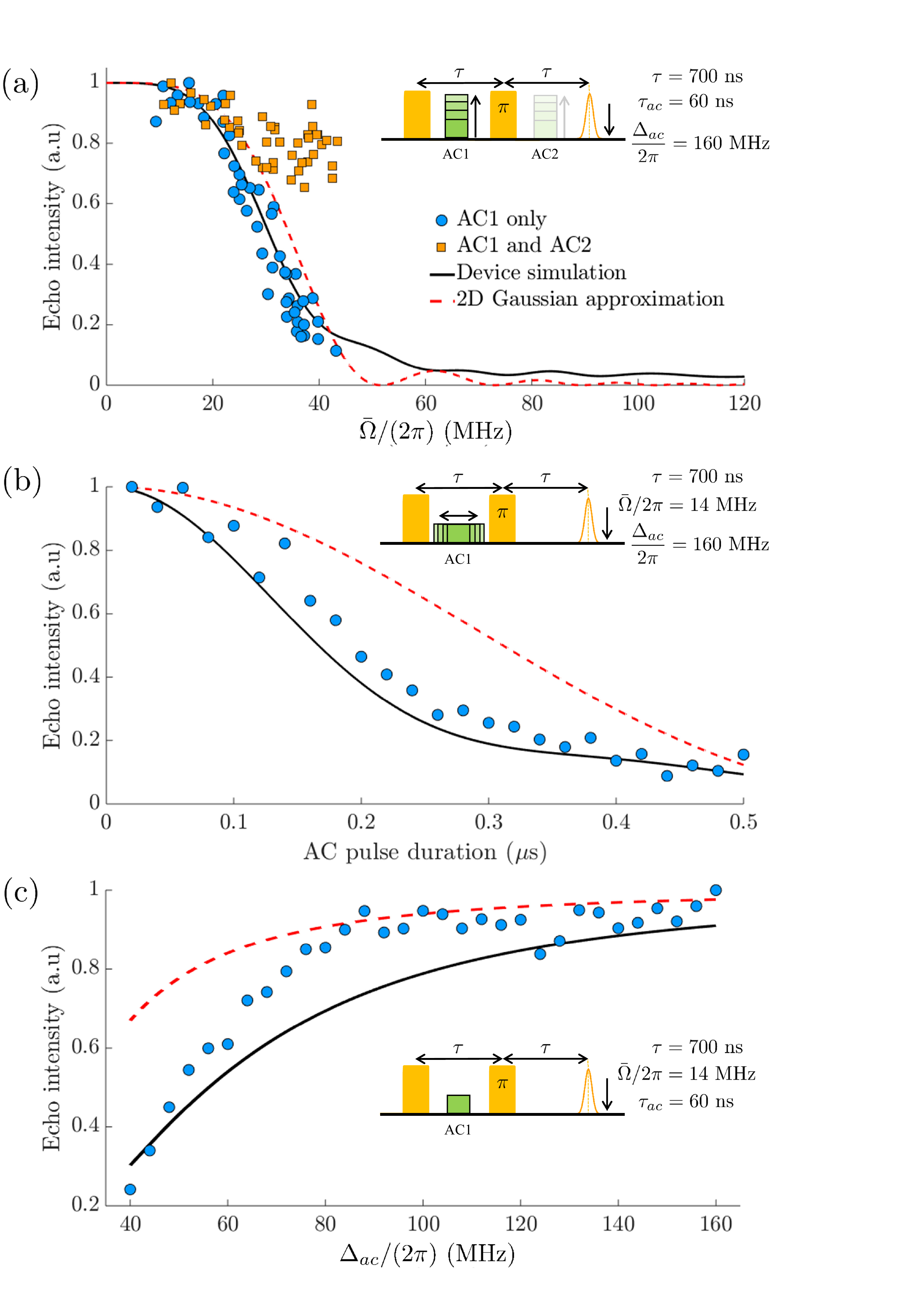}
  \end{center}
\caption{ACSS control of Nd\3\ photon echo emission. The echo intensity is plotted against the average ACSS Rabi frequency $\bar{\Omega}$ in (a), the ACSS duration $\tau_{ac}$ in (b), and the ACSS detuning $\Delta_{ac}$ in (c). The insets in each sub-figure illustrate the pulse sequence used. The duration of the input and $\pi$ pulses in all measurements was 20~ns, and $\tau$ represents the pulse-center to pulse-center time separation. The expected echo intensity based on the simulated cavity mode is shown by the solid black curve, and the analytic approximation derived from a 2d Gaussian distribution of the cavity field is shown by the dashed red curve. }
%We note that the calculated ratio of $\Omega(\bs{r})_{max}$ to the average Rabi frequency $\bar{\Omega}$ is 2.74 (see the Supplemental Material for further details).
\label{fig::Results}
\end{figure}

The ACSS is characterized through the study of two pulse photon echoes~\cite{Chaneliere2015, McDonald2016}, which were detected by photon counting with a silicon avalanche photodiode (APD). The two pulse echo sequence was augmented by additional off-resonant ACSS pulses (AC1 and AC2) before and after the inverting $\pi$-pulse (see the insets of Figure~\ref{fig::Results}). During the off-resonant pulses of length $\tau_{ac}$, the optical transition of each ion is frequency shifted by \mbox{$\delta_{ac}(\bs{r})\approx {\Omega(\bs{r})}^2 / (2 \Delta_{ac})$}, where $\Omega(\bs{r})$ is the Rabi frequency at spatial position $\bs{r}$, and $\Delta_{ac}$ is the detuning of the ACSS pulse from the echo input pulse. The resulting phase accumulated by each ion $\phi(\bs{r}) = \delta_{ac}(\bs{r}) \tau_{ac}$, is governed by the field amplitude of the cavity mode at the ion's spatial location. Because there is no correlation between an ion's resonant frequency and its position in the cavity, the application of an ACSS pulse results in an inhomogeneous phase shift across the ensemble. The inhomogeneity resulting from the ACSS pulse cannot be rephased by the optical $\pi$ pulse leading to a modulation of the photon echo intensity. However, rephasing the ACSS-induced inhomogeneity is possible through the application of additional ACSS pulses. To reduce any resonant interactions between the ions and the ACSS pulses, spectral trenches were prepared at the ACSS frequency prior to the sequence by optically pumping to the other electron spin level (see Figure~\ref{fig::DeviceAndSpectrum}(b)).     

Figure~\ref{fig::Results} shows the normalized intensity of the emitted photon echo for sequences that vary (a) the average Rabi frequency of the ACSS pulses $\bar{\Omega}$, (b) their duration, and (c) their relative frequency detuning. The data highlights the ability to control the coherent emission intensity by manipulating the relative phase evolution throughout the ensemble using the ACSS. In the case where only a single ACSS pulse is applied (solid blue circles in Figure~\ref{fig::Results}), the ACSS-induced inhomogeneity cannot be rephased and the echo is attenuated. The phase accumulated due to AC1 can  be balanced through the application of an identical ACSS pulse after the $\pi$-pulse, which in principle can restore the echo to full intensity. Figure~\ref{fig::Results}(a) (solid squares) demonstrates where this has been partially achieved through the application of AC2. On average the echo is restored to greater than 75\% of the unperturbed echo intensity~\footnote{In the work of Reference~\onlinecite{Chaneliere2015}, balancing ACSS pulses are able to recover the echo intensity with an efficiency of 98\%}. The incomplete recovery is likely to be dominated by imperfections in balancing the phase evolution from pulses AC1 and AC2, largely because of limitations in the timing resolution and intensity control of the applied ACSS pulses in our experimental setup (see further analysis in the Supplemental Material). Importantly, the restoration of the echo is evidence that the attenuation is caused by the ACSS interaction rather than by other dephasing processes such as instantaneous spectral diffusion, or device heating.

The echo intensity can be simulated using the cavity Maxwell-Bloch equations under the assumption of a uniform distribution of ions within the mode profile of the cavity (see Supplemental Material). Figure~\ref{fig::Results} shows normalized echo intensities from both a simplified analytical model (dashed line) and a numerical simulation (solid line). The analytical model assumed a two-dimensional Gaussian distribution~\cite{McDonald2016} for the cavity mode profile. This is a coarse approximation that captures the small variation of the field envelope along the y-axis in comparison to the variation along the x and z axes. Despite this, the analytical solution is a useful reference point for understanding the echo behavior. The numerical model used the simulated cavity mode profile from a finite difference time domain calculation (COMSOL), and the ions' frequencies are chosen from a Gaussian distribution with a FWHM equal to the input pulse Rabi frequency ($\bar{\Omega} = 2\pi\times25$~MHz). The numerical simulation is fitted to the data using one free parameter $R$: the ratio of the maximum ACSS Rabi frequency $\Omega(\bs{r})_{max}$ to the average ACSS Rabi frequency $\bar{\Omega}$. The agreement between the experimental and simulated data for the least squares fit value of $R = 1.83\pm0.02$ is further evidence that the ACSS is the dominant perturbation to the system.

With the experimentally determined value for $R$ it is possible to calculate the single photon-ion interaction strength $g$ by calibrating the average cavity photon population $\langle n \rangle$ to $\bar{\Omega}$. From the known transmission losses in our system we estimate that  $\langle n \rangle = 0.53$ for pulses with $\bar{\Omega} = 2\pi\times25$~MHz. The specified values of $R$ and $\langle n \rangle$ result in a $g = 2\pi\times31.4$~MHz. This is consistent with previous observations in this device~\cite{Zhong2017a} and only $\approx$10\% greater than the value of $g$ calculated from the published optical oscillator strength~\cite{Sun2002} and the simulated mode volume of the nanophotonic cavity (see Supplemental Material). Using the experimentally determined value of $g$, the maximum possible single photon ACSS in the cavity at a detuning $\Delta_{ac}/2\pi = 160$~MHz is $2g^2/\Delta_{ac} = 2\pi\times12.3$~MHz.

The results demonstrate that a single intra-cavity photon can produce an ACSS $\delta_{ac}$ that is $100\times$ larger than the Nd\3\ homogeneous linewidth $\Gamma_{\textrm{h}} \approx 100$~kHz (see Supplemental Material). Therefore, maintaining $\langle n \rangle \gg 0.01$ during $\tau_{ac}$ allows all-optical control of the relative phases of ions within an ensemble using time domain techniques, such as the photon echo. Thus, this work establishes a path toward realizing all-optical variations of techniques that have previously relied on applied electric fields~\cite{Chaneliere2015, McDonald2016}. 

In particular, our measurements form the basis for achieving an all-optical quantum memory based on the hybrid photon echo rephasing (HYPER) protocol previously implemented with electric field gradients~\cite{McAuslan2011a}. The HYPER protocol uses two inversion pulses to recall an input photon (the HYPER echo) when the ensemble is almost completely in the ground state. This avoids the stimulated emission noise that occurs when the recalled photon is emitted whilst the ensemble is inverted, such as in the case when only a single inversion pulse is used~\cite{Ruggiero2009}. For HYPER to achieve high efficiency, the intermediate echo resulting from the first inversion is suppressed using a controlled phase perturbation throughout the ensemble, which is later balanced to recover the HYPER echo. The largest benefit of HYPER is that, in the ideal implementation, no preparation of the inhomogeneous line is required, allowing the full optical depth and natural bandwidth of the material to be harnessed for quantum optical storage.

\begin{figure}[t]
  \begin{center}
\includegraphics[width=\columnwidth, trim = 6cm 6.5cm 7.5cm 6.5cm, clip = true]{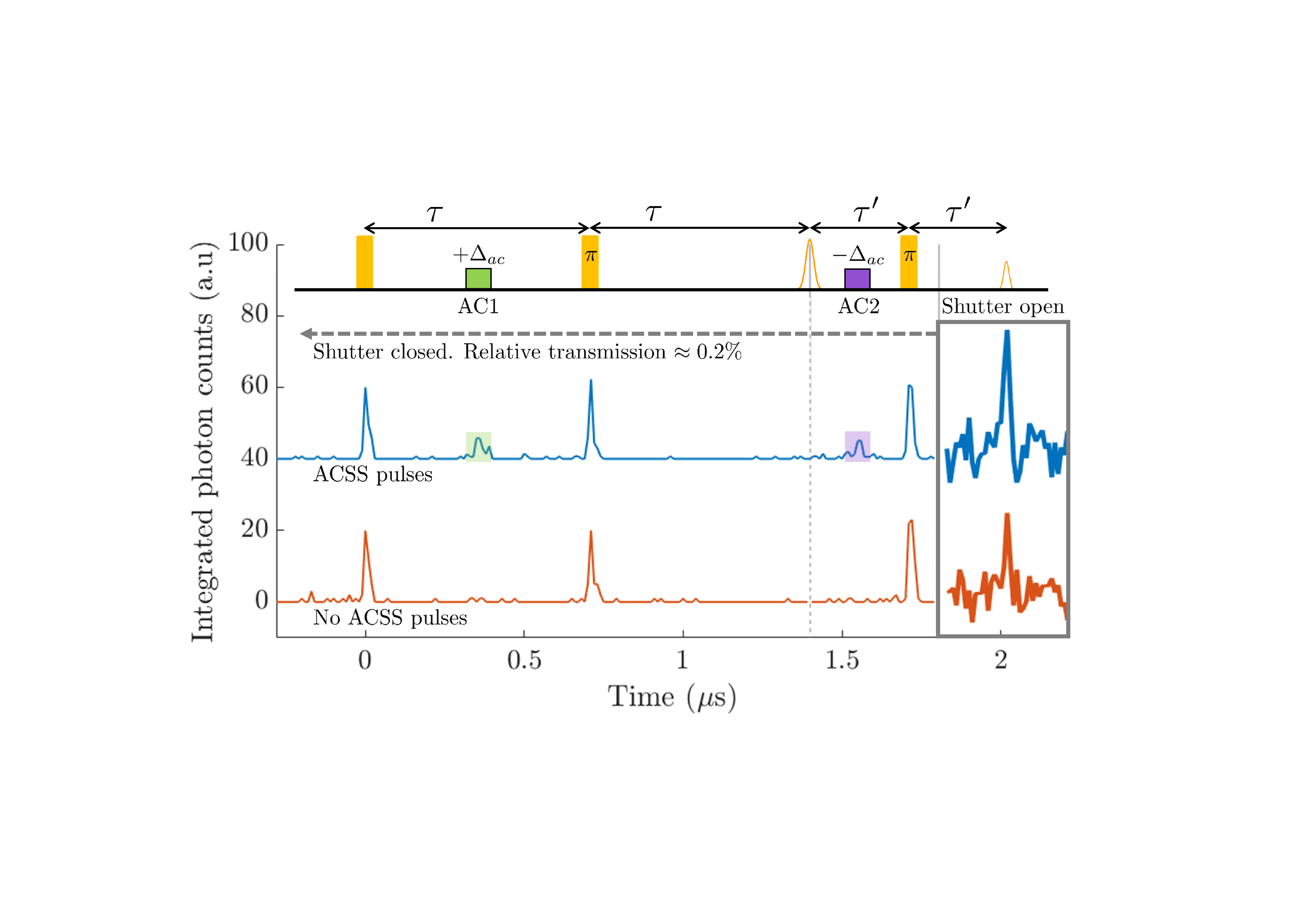}
  \end{center}
\caption{ACSS controlled HYPER protocol sequence. The lower trace shows the signal when no ACSS pulses are applied, and the upper trace (offset for clarity) shows the signal when both ACSS pulses are applied. Superimposed above the data is the pulse sequence used, where $\tau = 700$~ns, $\tau^{\prime} = 250$~ns, and $\Delta_{ac} = 160$~MHz. Further details are provided in the Supplemental Material.}
\label{fig::HYPER}
\end{figure}

The photon echo measurements presented in Figure~\ref{fig::Results}, demonstrate two of the important aspects for an all-optical HYPER memory. The first is the suppression of the intermediate echo using the controlled phase perturbation, and the second is the balancing of that phase to allow the formation of the HYPER echo (Figure~\ref{fig::Results}(a)). Both of these aspects are combined in a proof-of-principle demonstration of the HYPER sequence shown in Figure~\ref{fig::HYPER}, where the secondary echo is enhanced when the balanced ACSS pulses are applied. An all-optical version of HYPER is promising for on-chip quantum memories because additional electrodes on the REI substrate are not required, the efficiency can theoretically approach unity in the limit of an impedance matched cavity (see Supplemental Material), and $\Delta_{ac}$ can be increased so that the storage bandwidth can approach the inhomogeneous linewidth (GHz). Achieving the impedance matching condition requires approximately doubling the nanocavity quality factor, which is possible through improved fabrication. A challenge for HYPER memories is to simultaneously achieve high efficiency and high fidelity. To do so requires efficient inversion pulses over the bandwidth of interest, which is not achieved by the simple Gaussian pulses used in this work. Although the use of more complex adiabatic pulses offer a pathway to achieve large bandwidth and efficient inversion~\cite{Rippe2005}, the resultant instantaneous spectral diffusion may ultimately reduce the maximum storage time of the memory~\cite{Dajczgewand2015}.

Whilst well suited to the current devices, the HYPER memory is not the only protocol that could be achieved all-optically in the nanodevices. This is because the $\delta_{ac}$ of a single photon is large compared to the width of absorption features prepared by optical pumping ($\Gamma_f \leq 3$~MHz~\cite{Afzelius2010, Zhong2017a}). As a result it is possible to create an atomic frequency comb (AFC) memory with an ACSS controlled readout delay. This was demonstrated in Reference~\onlinecite{Zhong2017a}, where ACSS pulses were applied frequency-symmetrically at large detunings about the AFC center. Importantly, the ACSS allows control that is not possible using dc electric fields. Although protocols that dynamically alter the comb profile during storage using dc electric fields have been investigated~\cite{Lauritzen2011}, they are not able to achieve a continuously tunable delay. This is because there is no correlation between an ions' spatial position and resonant frequency in stochastically doped crystals. Because the ACSS is spectrally dependent, it is possible to achieve a continuously tunable storage time (see Supplemental Material). To realize ACSS controlled AFC memories operating with high efficiency further steps are required. In samples with a uniform distribution of ions the recall efficiency of a pulse stored with a controlled delay will be limited due to the ACSS inhomogeneity~\cite{Zhong2017a} (see Supplementary Material). To overcome this limitation requires control of the spatial location of the interacting ions within the cavity either through spectroscopic selection or controlled placement~\cite{Zhong2017a}. 

For both the ACSS controlled HYPER and AFC protocols, operation at the quantum level will require the suppression of noise photons that are generated by ions excited resonantly or off-resonantly by the ACSS control pulses. Therefore, a large single photon ACSS is desirable because the required frequency shift can be achieved with fewer photons, reducing the number of excited ions contributing to the noise. To suppress noise photons that are generated outside the memory bandwidth with high extinction, spectral filters created by optical pumping in another Nd\3:\YVO\ crystal can be applied~\cite{Beavan2013, Kutluer2017}. For photons generated within the memory bandwidth, a high memory efficiency ensures that these noise photons are time-separated from the signal photon through the protocol storage (see Supplemental Material for quantitative analysis using the AFC as an example). 

In addition to offering opportunities for advancing quantum memory protocols, large single photon ac Stark interactions can be harnessed for quantum non-demolition measurements~\cite{Sinclair2016}. Sinclair \etal\ discuss the measurement of the phase shift of an optical probe pulse stored in an AFC due to the ACSS of a single photon transmitted through the cavity in a transparent window adjacent to the memory. The phase shift of the retrieved probe pulse is then a non-destructive measurement that heralds the presence or absence of a single photon. The maximum single photon phase shift resultant from the experiments performed here is $3\times10^{-4}$~rad, which is consistent with the prediction from Reference~\onlinecite{Sinclair2016}. The parameters used in this work would require the probe pulse to be stored in a comb with a bandwidth of the order of 10s of MHz. Ideally, the phase shift would be increased further through longer optical confinement in the cavity. Given the current quality factor of the device ($Q \approx 2.8\times10^3$), an order of magnitude improvement should be possible with further optimization of the cavity design and fabrication, and would not require significant changes to the proposed scheme. 

Although an accumulated phase shift can be increased by improving the cavity $Q$, increasing the single photon interaction strength (the single ion $g$) requires a significantly lower cavity mode volume or an optical transition with a larger dipole moment. The photon-ion interaction strength achieved in our current generation of nanocavities approaches the limit for what can be achieved for REI quantum devices using a conventional photonic crystal structure. This is because the optical dipole moment of Nd\3:\YVO\ is among the largest for 4f \myto\ 4f transitions, and the mode volume of the device studied in this work is within a factor of 10 of the minimum mode volume for conventional dielectric photonic crystal cavities $\approx (\lambda/2n)^3$~[\onlinecite{Coccioli1998}].  

A relevant next goal is to realize a single photon ACSS that is large compared to the ensemble inhomogeneous linewidth, which is of the order of 100 MHz - 1 GHz. This would allow a single photon to create a cross phase shift approaching $10^{-2}$~rad, facilitating single shot, non-destructive quantum measurement, or to tune two ions into resonance with one another. To achieve a single photon ACSS of this order would require an increase in $g$ by a factor between 3 and 10 (the ACSS is proportional to $g^2$). One effective strategy to pursue this goal would be to change the cavity design to incorporate dielectric discontinuities~\cite{Robinson2005, Hu2016, Choi2017}, thereby reducing the mode volume. Previous work in the design of such cavities indicate that a 100$\times$ reduction in the mode volume is certainly feasible. A second strategy would be to use the weaker 4f \myto\ 4f transition for the storage of photonic qubits and the allowed 4f \myto\ 5d transitions to perform the ACSS manipulation. The optical dipole moments of the 4f \myto\ 5d transitions are of the order of 50$\times$ larger than the parity disallowed 4f \myto\ 4f transitions, potentially increasing the ac Stark interaction by a factor of over $10^3$. Although doubly resonant photonic crystal cavity designs exist~\cite{Rivoire2011}, integrating two cavities, one of which is required to operate close to the edge of the ultraviolet A band~\cite{Neel2014}, would be challenging. Currently, the most direct path toward such a device would use the hybrid approach that confines the light in a device layer bonded to an active REI substrate~\cite{Marzban2015, Miyazono2017}.  

In this paper we have demonstrated and characterized a large single photon ACSS of Nd\3 ions in a nanophotonic resonator fabricated on a \YVO\ substrate. By combining the relatively large optical dipole moment of the studied transition with the high spatial mode confinement of the on-chip photonic crystal cavity, it is possible to access a new regime for all-optical control in REI crystals. Importantly, the ACSS due to a single photon 160~MHz off-resonance was large compared to both the homogeneous line width and spectral feature width measured in this experiment configuration. Consequently, new opportunities arise for using single photons to control quantum information protocols including memory and detection schemes in this class of materials. Given the results of this work and the avenues for increasing the strength of the interaction, the ACSS is able to extend the versatility of an already appealing physical system for photon-spin integration at the quantum level.

\acknowledgments{}
This work was funded by a National Science Foundation (NSF) Faculty Early Career Development Program (CAREER) award (1454607), the AFOSR Quantum Transduction Multidisciplinary University Research Initiative (FA9550-15-1-002), and the Defense Advanced Research Projects Agency Quiness program (W31P4Q-15-1-0012). Equipment funding was also provided by the Institute of Quantum Information and Matter, an NSF Physics Frontiers Center with support from the Moore Foundation. The device nanofabrication was performed in the Kavli Nanoscience Institute at the California Institute of Technology. J.G.B. acknowledges the support of the American Australian Association’s Northrop Grumman Fellowship.

\bibliographystyle{aipnum4-1}
\bibliography{ACSS_library}

\end{document}